\documentclass[journal]{IEEEtran}
\usepackage{cite}
\usepackage{multirow}
\usepackage{multicol}

\usepackage[open]{bookmark}

\usepackage{threeparttable}
\usepackage{booktabs}
\usepackage{array}
\usepackage{makecell}

\ifCLASSINFOpdf
  \usepackage[pdftex]{graphicx}
\else
\fi
\usepackage{amsmath}
\newcommand{\norm}[1]{\left\lVert#1\right\rVert}

\usepackage{algorithmic}
\usepackage{amsmath,amssymb,amsfonts,dsfont}
\usepackage{algorithm, algorithmic}

\usepackage{array}

\usepackage{xcolor}

\usepackage{url}

\begin{document}
%
\title{Cross-domain Joint Dictionary Learning \\for ECG Inference from PPG}
%
%
%

\author{Xin~Tian,~\IEEEmembership{Student Member,~IEEE,}
        Qiang~Zhu,~\IEEEmembership{Student Member,~IEEE,}
        Yuenan~Li,~\IEEEmembership{Member,~IEEE,}
        Min~Wu,~\IEEEmembership{Fellow,~IEEE}
\thanks{X. Tian, Q. Zhu, Y. Li and M. Wu are with the Department of Electrical and Computer Engineering, University of Maryland, Collge Park, MD, 20742 USA (e-mail: {xtian17, zhuqiang, minwu}@umd.edu.)}
\thanks{Y. Li is also with the School of Electrical and Information Engineering. Tianjin University, Tianjin, China (e-mail: ynli@tju.edu.cn)}}

\maketitle

\begin{abstract}
The inverse problem of inferring electrocardiogram (ECG) from photoplethysmogram (PPG) is an emerging research direction that combines the easy measurability of PPG and the rich clinical knowledge of ECG for long-term continuous cardiac monitoring. The prior art for reconstruction using a universal basis has limited fidelity for uncommon ECG waveform shapes due to the lack of rich representative power. In this paper, we design two dictionary learning frameworks, the cross-domain joint dictionary learning (XDJDL) and the label-consistent XDJDL (LC-XDJDL), to further improve the ECG inference quality and enrich the PPG-based diagnosis knowledge. Building on the K-SVD technique, our proposed joint dictionary learning frameworks aim to maximize the expressive power by optimizing simultaneously a pair of signal dictionaries for PPG and ECG with the transforms to relate their sparse codes and disease information. The proposed models are evaluated with 34,000+ ECG/PPG cycle pairs containing a variety of ECG morphologies and cardiovascular diseases. We demonstrate both visually and quantitatively that our proposed frameworks can achieve better inference performance than previous methods, suggesting an encouraging potential for ECG screening using PPG based on the proactive learned PPG-ECG relationship.

\end{abstract}


%
\IEEEpeerreviewmaketitle

\section{Introduction}

\IEEEPARstart{C}{ardiovascular} diseases (CVDs) have become a leading cause of death globally. From alarming reports of the World Health Organization, an estimated 17.9 million people died from CVDs in 2016, representing $31\%$ of all global deaths \cite{cvd}.  

Electrocardiogram (ECG) is a widely-used gold-standard for the cardiovascular diagnostic procedure. By measuring the electrical activity of the heart and conveying information regarding heart functionality, continuous ECG monitoring is proven to be beneficial for early detection of CVDs~\cite{rosiek2016risk}. However, most conventional ECG equipment is restrictive on users' activities. Newer clinical ambulatory ECG monitoring devices, such as the Zio patch \cite{zio}, have alleviated the above-mentioned issues, although potential skin irritation during long-term adhesive wear remains, especially for people with sensitive skin. Apple Watch and wearable devices alike can show real-time ECG without adhesion to the skin, but generally requires active user participation and is usually for short duration measurement, making it infeasible for long-term continuous ECG monitoring.

Given the constraints of the ECG sensors, researchers have made efforts towards the long-term continuous ECG monitoring by inferring ECG from optical sensors, such as the photoplethysmogram (PPG) sensors ubiquitously seen in the wearable devices~\cite{2019arXiv190410481Z, tian2020cross}. PPG has become a common modality for monitoring heart condition due to the maturity of the technology and low cost. It measures the optical response of the blood volume changes at the peripheral ends, including ear lobes, wrists, and fingertips~\cite{allen2007photoplethysmography}, and provides valuable information about the cardiovascular system via daily use of wrist watch or fingertip pulse oximeter. Compared to ECG, PPG is more user-friendly in long-term continuous monitoring without constant user participation. 

PPG and ECG are physiologically related as they embody the same cardiac process in two different signal sensing domains. The peripheral blood volume change recorded by PPG is influenced by the contraction and relaxation of the heart muscles, which are controlled by the cardiac electrical signals triggered by the sinoatrial node~\cite{joshi2014review}. The waveform shape (i.e. signal morphology), pulse interval, and amplitude characteristics of PPG provide important information about the cardiovascular system~\cite{allen2007photoplethysmography}, including heart rate, heart rate variability~\cite{gil2010photoplethysmography}, respiration~\cite{johansson2003neural}, and blood pressure~\cite{chua2010towards}. It would be beneficial to consider the inverse problem of inferring the medical gold-standard ECG signal using the PPG sensor, which facilitates home-centered personal healthcare, especially during the global health crisis.  

The previous reconstruction work~\cite{2019arXiv190410481Z} using a universal discrete cosine transform (DCT) basis has limited fidelity for uncommon ECG waveform shapes due to the lack of rich representative power, especially for the subject-independent case with a broader range of signal morphologies~\cite{yang2012coupled}. To address this, our work focused on a more versatile and adaptive framework based on dictionary learning to demonstrate the feasibility of ECG waveform inference from artifact-free PPG signal as an inverse filtering problem. For practical applications, we recognize the need for removing the motion artifacts from PPG in the ambulatory condition~\cite{zhang2014troika,schack2015new}. Our on-going proof-of-concept work has shown a promising synergy by incorporating denoising to remove PPG artifacts before carrying out the ECG inference, although we will not elaborate here given the focused scope of the current paper.

\begin{figure*}[!t]
\centering
\includegraphics[width = 6.8in]{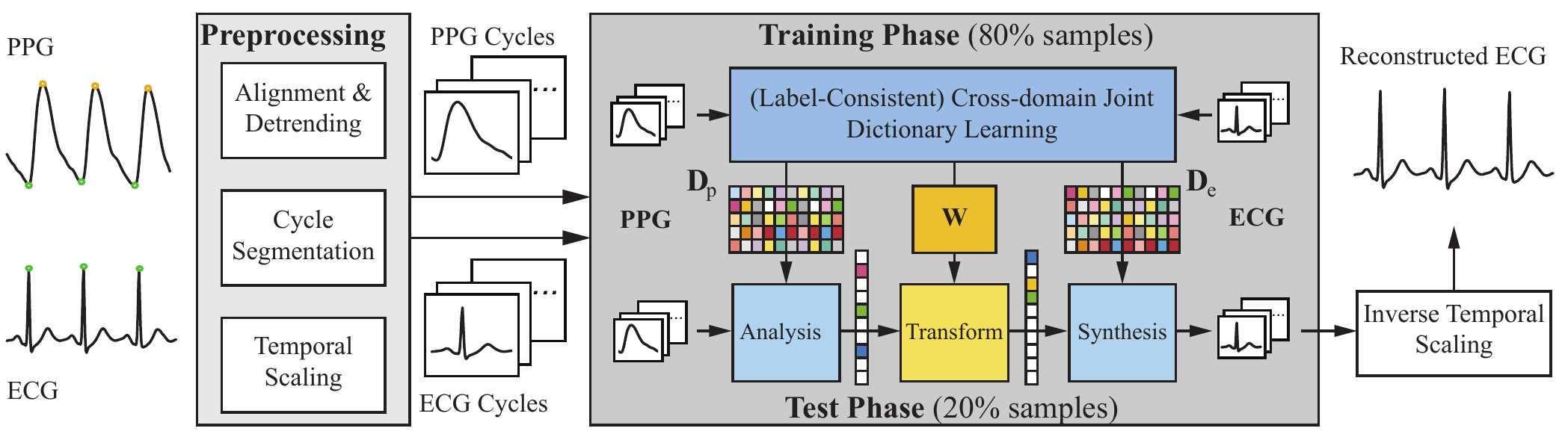}
\caption{Flowchart of the proposed framework. The ECG and PPG signals are first preprocessed to obtain temporally aligned and normalized pairs of cycles. $80\%$ pairs of ECG and PPG signal cycles from each subject are used for training paired dictionaries $\mathbf{D}_p$, $\mathbf{D}_e$, and a linear transform $\mathbf{W}$ which will be applied in the test phase to infer the ECG signals.}
\label{fig::sysdiag}
\end{figure*}

Our proposed cross-domain joint dictionary learning (XDJDL) method for ECG reconstruction from PPG is summarized in Fig.~\ref{fig::sysdiag}. A further-developed label-consistent XDJDL model (LC-XDJDL) is also proposed when the label information for the ECG/PPG cycles is available. The PPG and ECG signals are first preprocessed into normalized signal cycles to facilitate the subsequent training. In the training phase, the ECG/PPG dictionary pair is jointly updated with a stable linear mapping that relates the sparse representations of the two measurements. In LC-XDJDL, one more linear mapping that enforces the label-consistency for the PPG sparse codes will be learned to further improve the ECG reconstruction performance and enrich the PPG diagnosis knowledge base.

The main contributions of this paper include:
\begin{itemize}
\item As a clinical application of dictionary learning, a cross-domain joint dictionary learning (XDJDL) framework is proposed for ECG waveform inference from PPG. Our framework enforces that the dictionary pair has a good representation capability and that the sparse representation of ECG and PPG from the same cardiac cycle are related by a stable linear mapping. By adopting K-SVD for optimization, the mapping function is updated jointly with the dictionary pair, enabling a transform-aware training procedure. The linear transform reveals the intrinsic relation between atoms of the dictionary pair, showing a potential for preliminary diagnosis.
\item The label-consistent XDJDL (LC-XDJDL) adds the label-consistency constraint for PPG sparse codes. The benefits of this regularization in LC-XDJDL are not limited to improving the quality of ECG reconstruction, but also making the PPG based heart disease diagnosis more interpretable. The transform that relates the sparse codes of PPG to disease labels can help to gain intuitive insights on how a certain disease can impact on PPG.
\item The proposed XDJDL and LC-XDJDL frameworks are evaluated on a subset of a real-world clinical dataset, the MIMIC-III~\cite{johnson2016mimic}. This sub-dataset contains 34,000+ ECG/PPG cycle pairs with a large variety of ECG morphological patterns collected from different cardiovascular pathologies. The experimental result shows that XDJDL outperforms the state-of-the-art technique with remarkable improvements. And the reconstruction performance can be further improved if we can leverage the label information for the ECG/PPG cycles by using the LC-XDJDL model.
\item This work enables the synergistic utilization of the advantages of PPG and ECG signals for better preventive healthcare. It shows the potential of providing a more economical, user-friendly, and long-term cardiac monitoring methodology with a rich knowledge base of the clinical gold-standard ECG.
  
\end{itemize}

\section{Related Work}\label{sec:related works}
\subsection{ECG reconstruction from PPG}
Very limited prior work has been devoted to the PPG-based ECG inference. A machine learning method is presented~\cite{{banerjee2014photoecg}} to estimate typical ECG parameters from extracted PPG features with $90\%$ accuracy. However, the estimation for a limited number of ECG parameters is insufficient to provide full ECG screening for the currently trained medical professionals to make further decisions. A pilot study~\cite{2019arXiv190410481Z} proposed to relate the waveforms of PPG and ECG in the discrete cosine transform (DCT) domain by a linear model. This method achieved a mean reconstruction correlation as 0.98 in subject-specific cases. There is still substantial room for improvement when extending to the subject-independent case where a universal mapping needs to be trained from a wider variety of ECG morphologies of multiple people. To address these above-mentioned issues, we consider dictionary learning based sparse representation for ECG and PPG as it provides a richer and more adaptive representation than the universal dictionary DCT, and use this as a foundation to developing a joint dictionary learning method for reconstruction~\cite{tian2020cross}. 
\subsection{Dictionary learning}
Algorithms that learn a single dictionary for signal representation have been well-studied~\cite{aharon2006k, engan1999method, mairal2009supervised}. They have been successfully applied to cardiac signal processing, including recent research showing that ECG signals can be well-represented as a sparse linear combination of atoms from an appropriately learned dictionary for such applications as ECG classification and compression~\cite{liu2016dictionary,majumdar2017robust,craven2016adaptive}. 

In the domain of image processing and computer vision, these single dictionary learning strategies have been extended to joint dictionary learning tasks. For image super-resolution~\cite{yang2010image,yang2012coupled,xu2014coupled}, coupled dictionary learning frameworks are proposed to learn a dictionary pair for low- and high-resolution image patches while enforcing the similarity of their sparse codes with respect to their dictionaries. One assumption from this model is that the transform matrix between the two sparse codes is an identity matrix. In person re-identification~\cite{li2018discriminative} and photo-to-sketch~\cite{wang2012semi} problems, a linear mapping between the codings of input and output images is introduced into the objective function for semi-coupled dictionary learning. In both training schemes, the updates of the mapping and dictionaries are separately done within each iteration, making the dictionary computation less aware of the signal transform. 

Our method aims at boosting reconstruction performance from PPG to ECG by using a joint dictionary learning framework. Unlike the super-resolution problem~\cite{yang2010image,yang2012coupled,xu2014coupled} where the input and output reside in the same signal domain and highly correlated, XDJDL introduces a PPG-to-ECG mapping, which spans the two sensing modalities with low waveform correlation, providing more flexibility and generalization for the two learned dictionaries. Different from~\cite{li2018discriminative,wang2012semi}, we update the linear transform and the dictionary in the same step, which can optimize the capability of the obtained dictionaries for both signal representation and transformation. This kind of transform-aware joint dictionary learning formulation is one of the major differences from other coupled dictionary learning frameworks. This framework can also be easily generalized to different constraints. For instance, in the proposed LC-XDJDL model, we add a label-consistency regularization term to the objective function of the XDJDL model, which encourages the transformed sparse codes from the same class to be similar.

\section{Proposed System}
\label{sec::proposed system}
\subsection{Preprocessing}\label{subsec::preproc}

The goal of preprocessing ECG and PPG signals during the training phase is to obtain temporally aligned and normalized pair of signals, so that the critical temporal features of both
waveforms are synchronized for learning and evaluation.

We have followed the procedures used in~\cite{2019arXiv190410481Z}. First, we align the ECG and PPG sequences according to the moment when the ventricles contract and the blood flows to the vessels, which corresponds to the R peaks of ECG and the onsets of PPG in the same cycle. Both the onset and R peaks are detected by the beat detection functions from the PhysioNet Cardiovascular Signal Toolbox~\cite{vest2018open}. Then we detrend the aligned signals by a second-order difference operator based algorithm~\cite{2019arXiv190410481Z}  to eliminate the baseline drift related to respiration, motion, vasomotor activity, and change in contact surface~\cite{allen2007photoplethysmography}. The detrended PPG and ECG signals are partitioned into cycles by the \textit{R2R}~\cite{2019arXiv190410481Z} segmentation scheme, where the partition points are the R peaks of the ECG signal. After the segmentation, each cycle is linearly interpolated to length $d$ to mitigate the influence of the heart rate variation. Finally, we normalize the amplitude of each cycle by subtracting the sample mean and dividing by the sample standard deviation. The preprocessed PPG and ECG signal cycles will be stored in data matrices $\mathbf{P}$ and $\mathbf{E}$, respectively.

\subsection{Cross-domain Joint Dictionary Learning (XDJDL)}
\label{subsec:: proposed model and optimization}
\subsubsection{The K-SVD Model}
As one of the most popular dictionary learning methods, K-SVD model is composed of two main optimization steps: linear sparse coding based on the current overcomplete dictionary with $k$ atoms, and updating the dictionary by SVD method~\cite{aharon2006k}.

Let $\mathbf{X} \in \mathbf{R}^{d\times n}$ be a set of input signals, with each column $\mathbf{x}_i$ being a training sample. K-SVD aims to solve the following $L_0$-norm constraint problem in Eq.~\eqref{eq::ksvd}:
\begin{equation}\label{eq::ksvd}
    \begin{aligned}
    \underset{\mathbf{D},\mathbf{A}}{\mathrm{min}}&\norm{\mathbf{X}-\mathbf{DA}}_F^2\\
    \mathrm{s.t.} &\norm{\mathbf{a}_{j}}_0 \leq t_0,~~j = 1,...,n.
    \end{aligned}
\end{equation}
where $\mathbf{D} \in \mathbb{R}^{d \times k}$ is the reconstructive dictionary with $k$ atoms; $\mathbf{A} \in \mathbb{R}^{k \times n}$ is the corresponding sparse codes of $\mathbf{X}$, with each column denoted as $\mathbf{a}_j$; and $t_0$ is a parameter for sparsity constraint. Frobenius norm is used to calculate the element-wise ${l_2}$-norm in the given matrix.

In this paper, we aim to tackle the ECG inference from PPG by learning a dictionary pair for ECG and PPG along with a linear transform between the sparse representations of the two signals. The reconstructive dictionary pair characterizes the two structural domains of the two biomedical signals, and the mapping function can reveal the intrinsic relationship between ECG and PPG signals in the sparse domain. We impose the linear mapping error as one regularization term in the objective function, then convert it to be a problem that can be optimized by the K-SVD dictionary learning method. The details of the model formulation and optimization algorithm are discussed in the following subsections.

\subsubsection{Proposed XDJDL Model}
We denote the PPG and ECG datasets as $\mathbf{P} = [\mathbf{X}_p, \mathbf{T}_p] \in \mathbb{R}^{d\times(n+m)}$ and $\mathbf{E} = [\mathbf{X}_e, \mathbf{T}_e] \in \mathbb{R}^{d\times(n+m)}$ respectively. Each column of $\mathbf{P}$ and $\mathbf{E}$ is denoted as $\mathbf{p}_i \in \mathbb{R}^{d\times 1}$ and $\mathbf{e}_i \in \mathbb{R}^{d\times 1}$, representing one PPG/ECG cycle during the same cardiac cycle. The goal is to learn the patterns (in terms of dictionaries, mappings, etc.) from the training data $\mathbf{X}_p \in \mathbb{R}^{d\times n}$ and $\mathbf{X}_e \in \mathbb{R}^{d\times n}$ to infer the test ECG dataset $\mathbf{T}_e \in \mathbb{R}^{d\times m}$ from PPG $\mathbf{T}_p \in \mathbb{R}^{d\times m}$.
 
We formulate our XDJDL framework as:
\begin{equation}\label{equ:: math model 1}
    \begin{aligned}
    \underset{\substack{\mathbf{D}_e,\mathbf{A}_e,\mathbf{D}_p, \mathbf{A}_p,\mathbf{W}}}{\mathrm{min}} &\norm{\mathbf{X}_e - \mathbf{D}_e\mathbf{A}_e}_F^2
     + \alpha \norm{\mathbf{X}_p - \mathbf{D}_p\mathbf{A}_p}_F^2\\
    & + \beta\norm{\mathbf{A}_e - \mathbf{W}\mathbf{A}_p}_F^2 \\
     \mathrm{s.t.} &\norm{\mathbf{a}_{p,j}}_0 \leq t_p, \mathrm{and} \norm{\mathbf{a}_{e,j}}_0 \leq t_e, ~~j = 1,...,n.
    \end{aligned}
\end{equation}
where $\mathbf{D}_p \in \mathbb{R}^{d\times k_p}$ and $\mathbf{D}_e \in \mathbb{R}^{d\times k_e}$ are dictionaries learned for $\mathbf{X}_p$ and $\mathbf{X}_e$, respectively; $\mathbf{A}_p \in \mathbb{R}^{k_p\times n }$ and $\mathbf{A}_e \in \mathbb{R}^{k_e\times n }$ are the corresponding sparse coding matrices related with the data matrices $\mathbf{X}_p, \mathbf{X}_e$ when $\mathbf{D}_p, \mathbf{D}_e$ are the current dictionaries. Each column of $\mathbf{A}_p$ and $\mathbf{A}_e$ is denoted as $\mathbf{a}_{p,j}$ and $\mathbf{a}_{e,j}$ with the sparsity upper bounded by $t_p$ and $t_e$. 

For the objective function in Eq.~\eqref{equ:: math model 1}, $\norm{\mathbf{X}_e - \mathbf{D}_e\mathbf{A}_e}_F^2$ and $\norm{\mathbf{X}_p - \mathbf{D}_p\mathbf{A}_p}_F^2$ are the data fidelity terms for ECG and PPG cycle sets, respectively. The term $\norm{\mathbf{A}_e - \mathbf{W}\mathbf{A}_p}_F^2$ represents the mapping error between the sparse coding coefficients of ECG and PPG signals, which enforces the transformed sparse codes of PPG to approximate that of ECG. Intuitively, we can enforce the two sparse representations for ECG and PPG from the same cycle to be the same and set the regularization term as $\norm{\mathbf{A}_e - \mathbf{A}_p}_F^2$. However, since ECG and PPG are from two different signal sensing modalities and the waveform difference between the two signals is significant, directly pushing their sparse representations to be similar could compromise the generalization of the two learned dictionaries.

From the formulation in the Eq.~\eqref{equ:: math model 1}, we can jointly learn the dictionaries for ECG and PPG datasets, which produce a good representation for each sample in the training set with strict sparsity constraints. Meanwhile, we learn the linear approximation of the transform that relates the sparse codes of PPG and ECG, and use it to entail the intrinsic relationship between certain PPG atoms and ECG atoms from their corresponding dictionaries.

\subsubsection{Optimization}\label{optimization}
Eq.~(\ref{equ:: math model 1}) can be rewritten as:
\begin{equation}\label{equ::math model 2}
\begin{aligned}
\underset{\substack{\mathbf{D}_e,\mathbf{A}_e,\mathbf{D}_p,\\ \mathbf{A}_p,\mathbf{W}}}{\mathrm{min}}&\norm{
    \begin{pmatrix}
\mathbf{X}_e\\
\sqrt{\alpha} \mathbf{X}_p\\
\mathbf{0}\\
\end{pmatrix} - 
\begin{pmatrix}
\mathbf{D}_e & \mathbf{0} \\
\mathbf{0} & \sqrt{\alpha} \mathbf{D}_p \\
-\sqrt{\beta} \mathbf{I} & \sqrt{\beta} \mathbf{W}
\end{pmatrix}
\begin{pmatrix}
\mathbf{A}_e\\
\mathbf{A}_p
\end{pmatrix}
}_F^2\\
 \mathrm{s.t.} & \norm{\mathbf{a}_{e,j}}_0 \leq t_e, \mathrm{and} \norm{\mathbf{a}_{p,j}}_0 \leq t_p, ~~j = 1,...,n.
\end{aligned}
\end{equation}
where $\mathbf{I}$ is an identity matrix and $\mathbf{0}$ is a zero matrix, with valid dimensions for matrix multiplication.

Let 
$\mathbf{X} \triangleq (\mathbf{X}_e , \sqrt{\alpha} \mathbf{X}_p, \mathbf{0})^\text{T} \in \mathbb{R}^{(2d + k_e)\times n }$, 
$\mathbf{D} \triangleq (\mathbf{D}_e,\mathbf{0}, -\sqrt{\beta} \mathbf{I}; \mathbf{0}, \sqrt{\alpha} \mathbf{D}_p , \sqrt{\beta} \mathbf{W})^\text{T}\in \mathbb{R}^{(2d+k_e)\times (k_e + k_p) }$, and 
$\mathbf{A} \triangleq (\mathbf{A}_e , \mathbf{A}_p)^\text{T} \in \mathbb{R}^{(k_e + k_p)\times n }$. 
The optimization of (\ref{equ::math model 2}) can be written as the following problem:
\begin{equation}
    \begin{aligned}
    \underset{\mathbf{D,A}}{\mathrm{min}}&\norm{\mathbf{X} - \mathbf{D}\mathbf{A}}_F^2,\\
    \mathrm{s.t.} &\norm{\mathbf{a}_{+,j}}_0 \leq t_e, \mathrm{and} \norm{\mathbf{a}_{-,j}}_0 \leq t_p, ~~j = 1,...,n.
    \end{aligned}
    \label{eqn::math model 3}
\end{equation}
where $\mathbf{a}_{*,j}$ represents for the column of $\mathbf{A}_*$, and $\mathbf{A}_+$ is defined as the first $k_e$ rows of sparse matrix $\mathbf{A}$ while $\mathbf{A}_-$ is the last $k_p$ rows of sparse matrix $\mathbf{A}$. The formulation in Eq.~\eqref{eqn::math model 3} is now similar to Eq.~\eqref{eq::ksvd}, suggesting that K-SVD can be adapted for this optimization. The difference is the local sparsity constraint, which will be addressed in the following optimization procedures.
\\

\textbf{Step 0: Initialization}
\\

To initialize $\mathbf{D}~\text{and}~\mathbf{A}$, we need to initialize their components: $\mathbf{D}_e,\mathbf{D}_p,\mathbf{W}, \mathbf{A}_e,$ and $\mathbf{A}_p$.  First, we randomly select a subset of columns from training data $\mathbf{X}_e$ and $\mathbf{X}_p$ to form $\mathbf{D}_e$ and $\mathbf{D}_p$. Then, we initialize the sparse codes $\mathbf{A}_e$ and $\mathbf{A}_p$ by solving Eq.~\eqref{eqn::sparse coding} with respect to \{$\mathbf{D}_e,~\mathbf{X}_e,~t_e$\} and \{$\mathbf{D}_p,~\mathbf{X}_p,~t_p$\}, respectively. Finally, we use the ridge regression model to initialize $\mathbf{W}$:

\begin{equation}
    \begin{aligned}
    \underset{\mathbf{W}}{\mathrm{min}} \norm{\mathbf{A}_e - \mathbf{W} \mathbf{A}_p}_F^2 + \lambda\norm{\mathbf{W}}_F^2.
    \end{aligned}
\end{equation}
 This has a closed-form solution as:
\begin{equation}
    \mathbf{W} = \mathbf{A}_e\mathbf{A}_p^\text{T}(\mathbf{A}_p\mathbf{A}_p^\text{T} + \lambda \mathbf{I})^{-1}.
    \label{eqn:ridge}
\end{equation}

After the initialization, we use a two-step iterative optimization to minimize the energy in (\ref{eqn::math model 3}), whereby step one is sparse coding and step two is dictionary updating by SVD.
\\

\textbf{Step 1: Sparse coding}
\\

Given D, the step of sparse coding finds the sparse representation $\mathbf{a}_j$ for $\mathbf{x}_j$, for $j = 1,...,n$, by solving
\begin{equation}
    \begin{aligned}
     \underset{\mathbf{a}_j}{\mathrm{min}}&\norm{\mathbf{x}_j - \mathbf{D}\mathbf{a}_j}_2^2   \\
    \mathrm{s.t.}& \norm{\mathbf{a}_{j}}_0 \leq t.
    \end{aligned}
    \label{eqn::sparse coding}
\end{equation}
where $\mathbf{a}_j$ is the $j^{th}$ column of the sparse representation matrix $\mathbf{A}$ and $\mathbf{x}_j$ is the $j^{th}$ training sample in matrix $\mathbf{X}$. 

Many approaches were proposed to solve Eq.~(\ref{eqn::sparse coding})~\cite{zhang2015survey}. Here we adopt the orthogonal matching pursuit (OMP) method\cite{tropp2007signal}, which is a greedy method that provides a good approximation. As mentioned earlier, the local sparsity constraints imposed on Eq.~\eqref{eqn::math model 3} will affect the direct application of OMP. One workaround is to solve the following problem in Eq.~\eqref{eqn::math model 4} in place of Eq.~\eqref{eqn::math model 3},  
\begin{equation}
    \begin{aligned}
    \underset{\mathbf{D,A}}{\mathrm{min}} & \norm{\mathbf{X} - \mathbf{D}\mathbf{A}}_F^2\\
    \mathrm{s.t.} & \norm{\mathbf{a}_{j}}_0 \leq t_e + t_p, ~~j = 1,...,n.
    \end{aligned}
    \label{eqn::math model 4}
\end{equation}
where $\mathbf{a}_j$ is the vertical concatenation of $\mathbf{a}_{+,j}$ and $\mathbf{a}_{-,j}$ in Eq.~\eqref{eqn::math model 3}, and $t_e$ and $t_p$ are the sparsity constraints for the upper and bottom parts of $\mathbf{a}_j$, respectively. During the OMP process in each iteration, we will only keep the largest sparse coefficients in $\mathbf{a}_j$ to ensure the local sparsity constraints.
\\

\textbf{Step 2: Dictionary update}
\\

To update the $k^{th}$ atom, $\mathbf{d}_k$, in dictionary $\mathbf{D}$ and its corresponding coefficients, $\mathbf{a}_R^k$, in the $k^{th}$ row of $\mathbf{A}$, we apply SVD to the residue term $\mathbf{R}_k \triangleq{\mathbf{X}-\sum_{j\neq k}\mathbf{d}_j\mathbf{a}_R^j}$. In practice, we only select the training samples that use the atom $\mathbf{d}_k$ and avoid filling in the zeros entries of $\mathbf{a}_R^k$ during the update. We do so through denoting the nonzero entries in $\mathbf{a}_R^k$ as $\mathbf{\tilde{a}}_R^k$, and correspondingly, $\mathbf{R}_k$ as $\mathbf{\tilde{R}}_k$. The updated atom $\mathbf{d}_k$ and the related coefficients $\mathbf{\tilde{a}}_R^k$ will then be computed by:
\begin{equation}
    \begin{aligned}
    \underset{\mathbf{d}_k,\mathbf{\tilde{a}}_R^k}{\mathrm{min}}\norm{\mathbf{\tilde{R}}_k-\mathbf{d}_k\mathbf{\tilde{a}}_R^k}^2_F.
    \end{aligned}\label{eqn::ksvd method}
\end{equation}

To solve Eq.~(\ref{eqn::ksvd method}), we use SVD method on the residue term~\cite{aharon2006k}, i.e. $\mathbf{\tilde{R}}_k = \mathbf{U\Sigma V^T}$. And then, $\mathbf{d}_k$ and $\mathbf{\tilde{a}}_R^k$ can be updated as follows:
\begin{equation}
    \begin{aligned}
    & \mathbf{d}_k = \mathbf{U}(:,1),\\
    & \mathbf{\tilde{a}}_R^k = \mathbf{\Sigma} (1,1) \mathbf{V^T}(1,:).
    \end{aligned}
    \label{eqn::sln to ksvd method}
\end{equation}

Note that taking $\mathbf{D} \triangleq (\mathbf{D}_e,\mathbf{0}, -\sqrt{\beta} \mathbf{I}; \mathbf{0}, \sqrt{\alpha} \mathbf{D}_p , \sqrt{\beta} \mathbf{W})^\text{T}$ as a whole in the dictionary update phase does not solve this optimization problem because the zero matrices part and the identity matrix part in $\mathbf{D}$ cannot be guaranteed in the update of the dictionary by SVD. A remedy to the above problem is to decompose the dictionary update problem for $\mathbf{D}$ into the following two subproblems by revisiting the matrix form of the optimization problem in Eq.~\eqref{equ::math model 2}.\\
(i) Update $\mathbf{D}_e, \mathbf{A}_e$:
\begin{equation}
    \begin{aligned}
    <\mathbf{D}_e^*,\mathbf{A}_e^*> = \underset{\mathbf{D}_e,\mathbf{A}_e}{\mathrm{argmin}}\norm{\mathbf{X}_e-\mathbf{D}_e\mathbf{A}_e}_F^2.\\
    \end{aligned}
    \label{eqn::subproblem1}
\end{equation}

We use SVD to update all atoms in $\mathbf{D}_e$ and the corresponding nonzero entries in $\mathbf{A}_e$ by solving Eq.~(\ref{eqn::subproblem1}) with the same procedure as in Eq.~(\ref{eqn::ksvd method}) and (\ref{eqn::sln to ksvd method}). The columns of $\mathbf{D}_e$ are $l_2$ normalized.\\
(ii) Update $\mathbf{D}_p, \mathbf{A}_p,$ and $\mathbf{W}$:

The updated ECG sparse representation matrix $\mathbf{A}_e^*$ from the subproblem (i) then serves as an input to the second subproblem here to update $\mathbf{W}$, $\mathbf{D}_p$ and $\mathbf{A}_p$ in Eq.~\eqref{eqn::subproblem2}.

\begin{small}
\begin{equation}
    \begin{aligned}
    < \mathbf{D}_p^*,\mathbf{A}_p^*,\mathbf{W}^*> =\underset{\mathbf{D}_p,\mathbf{A}_p,\mathbf{W}}{\mathrm{argmin}}\norm
    {
    \begin{pmatrix}
    \sqrt{\alpha} \mathbf{X}_p \\
    \sqrt{\beta} \mathbf{A}_e^*
    \end{pmatrix}
    -
    \begin{pmatrix}
    \sqrt{\alpha} \mathbf{D}_p \\
    \sqrt{\beta} \mathbf{W}
    \end{pmatrix}
    \mathbf{A}_p
    }_F^2.
    \end{aligned}
    \label{eqn::subproblem2}
\end{equation}
\end{small}

We treat $(\sqrt{\alpha} \mathbf{D}_p,\sqrt{\beta} \mathbf{W})^\text{T}$ as a whole dictionary, and use the SVD method in Eq. \eqref{eqn::ksvd method} and \eqref{eqn::sln to ksvd method} to update it together with the nonzero entries in $\mathbf{A}_p$. The linear transform and the dictionary are updated simultaneously, which addresses the problem of isolated update raised in~\cite{wang2012semi,li2018discriminative} and is one of the major differences from other coupled dictionary learning models.

After solving the two subproblems, $\mathbf{D}~\text{and}~\mathbf{A}$ can be assembled by filling in the submatrices. The main steps of XDJDL are summarized in Algorithm~\ref{algo1}. 

\begin{algorithm}[ht]

\caption{Cross-domain joint dictionary learning}
\label{algo1}

\begin{algorithmic}

\STATE{\textbf{Input:} Training data $\mathbf{X}_e$ and $\mathbf{X}_p$ of ECG and PPG cycles, Testing data $\mathbf{T}_e$ and $\mathbf{T}_p$, and sparsity constraints $t_e,~t_p$}

\hrulefill

\STATE{\textit{Training phase:}}
\STATE{\textbf{Initialization:}
\begin{itemize}
    \item Initialize $\{\mathbf{D}_e,\mathbf{D}_p\}$ by randomly selecting atoms from the training data.
    \item Initialize $\mathbf{A}_e,\mathbf{A}_p$ by solving Eq.~(\ref{eqn::sparse coding}) with OMP.
    \item Initialize $\mathbf{W}$ by Eq.~(\ref{eqn:ridge}).
\end{itemize}}

\WHILE{not converged}
\STATE{
\begin{itemize}
    \item{Update $\mathbf{D}, \mathbf{A}$ by combining updated submatrices.}
    \item{Sparse coding: compute $\mathbf{A}$ in Eq.~\eqref{eqn::sparse coding} with OMP. Zero out the smallest nonzero entries in the columns of $\mathbf{A}$ if any local sparsity constraint does not hold.
    \item{Dictionary update: 
    \begin{itemize}
        \item Update $\mathbf{D}_e, \mathbf{A}_e$ in Eq.~(\ref{eqn::subproblem1}) by SVD method illustrated in Eq.~(\ref{eqn::ksvd method})(\ref{eqn::sln to ksvd method}).
        \item Update $\mathbf{D}_p, \mathbf{A}_p, \mathbf{W}$ in Eq.~(\ref{eqn::subproblem2}) by SVD method illustrated in Eq.~(\ref{eqn::ksvd method})(\ref{eqn::sln to ksvd method}).
    \end{itemize}}
}
\end{itemize}}

\ENDWHILE

\STATE{\textit{Testing phase:}}
\FOR{each sample $\mathbf{t}_p^j \in \mathbf{T}_p$}
\STATE{
\begin{itemize}
    \item Compute sparse code $\mathbf{s}_p^j$ of $\mathbf{t}_p^j$ under $\mathbf{D}_p$ using Eq.~(\ref{eqn::sparse coding}).
    \item Calculate $\mathbf{s}_e^j = \mathbf{W}\mathbf{s}_p^j$.
    \item Compute the reconstructed ECG sample as $\mathbf{r}_e^j = \mathbf{D}_e\mathbf{s}_e^j$, and store it in matrix $\mathbf{R}_e$.
\end{itemize}
}
\ENDFOR\\
\hrulefill
\STATE{\textbf{Output: $\mathbf{R}_e$}}
\end{algorithmic}

\end{algorithm}



\subsection{Label Consistent XDJDL (LC-XDJDL)}

\label{lc-XDJDL}
For cases where the disease type is known or can be predicted, such as from the PPG signals that we have, we can further leverage the disease label. In this subsection, we examine the effect of adding a label consistency regularization term to the objective function in Eq.~\eqref{equ:: math model 1} as follows:
\begin{equation}\label{equ:: math model LC-XDJDL}
    \begin{aligned}
    \underset{\substack{\mathbf{D}_e,\mathbf{A}_e,\mathbf{D}_p, \mathbf{A}_p,\mathbf{W}}}{\mathrm{min}} &\norm{\mathbf{X}_e - \mathbf{D}_e\mathbf{A}_e}_F^2
     + \alpha \norm{\mathbf{X}_p - \mathbf{D}_p\mathbf{A}_p}_F^2\\
    & + \beta\norm{\mathbf{A}_e - \mathbf{W}\mathbf{A}_p}_F^2 + \gamma \norm{\mathbf{Q} - \mathbf{H}\mathbf{A}_p}_F^2\\
     \mathrm{s.t.} &\norm{\mathbf{a}_{p,j}}_0 \leq t_p, \mathrm{and} \norm{\mathbf{a}_{e,j}}_0 \leq t_e, ~~j = 1,...,n.
    \end{aligned}
\end{equation}
where $\mathbf{Q} \triangleq [q_1,q_2,..., q_n] \in \mathbb{R}^{r\times n}$ is a discriminative representation matrix~\cite{jiang2013label} in which each column $q_i = [0,0,...,0,1,1,0,..,0]^\text{T} \in \mathbb{R}^{r \times 1}$ corresponds to a discriminative coding for an input signal. The nonzero elements in $q_i$ occur at the corresponding disease label, which is similar to the one-hot encoding with the number of ones as a tunable parameter. 
The additional regularization term $\norm{\mathbf{Q} - \mathbf{H}\mathbf{A}_p}_F^2$ represents the discriminative sparse code error, which enforces the transformed sparse codes of PPG to approximate the discriminative codes in $\mathbf{Q}$. It yields such dictionaries that the signals from the same class have very similar sparse codes, i.e. enforcing the label-consistency in the sparse representations.

We add the label-consistency regularization term for two main purposes: One is to improve the ECG reconstruction quality by using additional class information to constrain the degrees of freedom of the PPG sparse codes. The other is to enrich the knowledge base of PPG for diagnosis of a certain set of diseases of interest. By inspecting the specific columns of $\mathbf{A}_p$ and $\mathbf{H}$, one can gain insights on how the disease is revealed on PPG.

Similarly, Eq.~ \eqref{equ:: math model LC-XDJDL} can be written in the matrix form:
\begin{equation}\label{equ::math model 2-LC-XDJDL}
\begin{aligned}
\underset{\substack{\mathbf{D}_e,\mathbf{A}_e,\mathbf{D}_p,\\ \mathbf{A}_p,\mathbf{W}}}{\mathrm{min}}&\norm{
    \begin{pmatrix}
\mathbf{X}_e\\
\sqrt{\alpha} \mathbf{X}_p\\
\mathbf{0}\\
\mathbf{Q}\\
\end{pmatrix} - 
\begin{pmatrix}
\mathbf{D}_e & \mathbf{0} \\
\mathbf{0} & \sqrt{\alpha} \mathbf{D}_p \\
-\sqrt{\beta} \mathbf{I} & \sqrt{\beta} \mathbf{W}\\
0 & \sqrt{\gamma}\mathbf{H}
\end{pmatrix}
\begin{pmatrix}
\mathbf{A}_e\\
\mathbf{A}_p
\end{pmatrix}
}_F^2\\
 \mathrm{s.t.} & \norm{\mathbf{a}_{e,j}}_0 \leq t_e, \mathrm{and} \norm{\mathbf{a}_{p,j}}_0 \leq t_p, ~~j = 1,...,n.
\end{aligned}
\end{equation}
The two-step optimization method in Section~\ref{optimization} can still be applied to find the optimal solution to both the dictionary pair and the linear mappings $\mathbf{W}$ and $\mathbf{H}$. In the test phase, the PPG sparse representation matrix $\mathbf{A}_p$ is obtained by applying sparse coding with the learned $\mathbf{D}_p$, $\mathbf{H}$, the test sample matrix $\mathbf{T}_p$, and the label matrix $\mathbf{Q}$.

\section{Experimental Evaluation}\label{sec:exper}
\subsection{Dataset} \label{subsec::dataset}

The Medical Information Mart for Intensive Care III (MIMIC-III)~\cite{johnson2016mimic, goldberger2000physiobank} is a publicly-available database assembled by researchers at the MIT. It comprises a large number of ICU patients with de-identified health data from their hospital stays. To evaluate our proposed framework and algorithm, we have extracted a small subset of the MIMIC-III database as follows. First, we select waveforms that contain both lead-II ECG and PPG signals sampled at 125Hz from the MIMIC-III waveform database. Then the selected waveforms are cross-referenced with the corresponding patient profile by subject ID in the MIMIC-III clinical information database. Patients with the four types of CVDs are further selected: congestive heart failure (CHF), myocardial infarction (MI) including ST-segment elevated (STEMI) and non-ST segment elevated (NSTEMI), hypotension (HYPO), and coronary artery disease (CAD). These diseases are all included in the ``diseases of the circulatory system" in the ICD-9 international disease classification codes. After that, we analyze the signal pair quality using the PPG SQI function from the PhysioNet cardiovascular signal toolbox~\cite{vest2018open} and keep the pair segments that are evaluated as ``acceptable" or ``excellent."

The resulting dataset consists of $33$ patients, with each patient having only one of the four diseases in the record. Each patient has three sessions of $5$-min ECG and PPG paired recordings collected within several hours, resulting in $34,000+$ ECG/PPG cycle pairs in total. Table~\ref{tab::mimic_class} shows the composition of the collected dataset.

\begin{table}[!t]
\centering
\caption{Composition of the collected dataset}
\label{tab::mimic_class}
\begin{tabular}{@{}lcc@{}}\toprule
Disease & Number of patients& Number of cycles\\
\hline
CHF& 7 & 7075 (20.6$\%$) \\ 
MI& 7 & 7106 (20.8$\%$)\\
HYPO& 7 & 8281 (24.2$\%$) \\
CAD& 12 & 11781 (34.4$\%$)\\ 
\hline
Total & 33 & 34243 (100$\%$)\\
\bottomrule
\end{tabular}
\begin{tablenotes}
 \scriptsize
 \item{~~~~CHF: congestive heart failure~~~~~~~~~~~~~~~~MI: myocardial infarction}
 \item{~~~~HYPO: hypotension~~~~~~~~~~~~~~~~~~~~~~~~~~~CAD: coronary artery disease}
\end{tablenotes}
\end{table}

\subsection{Metrics for Evaluation}

As shown in Fig.~\ref{fig::wave seg}~(a), a complete ECG cycle contains five major points, including P, Q, R, S, and T, which segment the ECG cycle into P wave, QRS complex, and T wave. The shape information of those waves is useful for further diagnosis. The interval parameters (PR interval, QRS interval, QT interval) defined by those five fiducial points are also important for examining a patient's heart conditions. Thus, to evaluate the quality of the reconstructed ECG, we take both into consideration the morphological metrics and the accuracy of time interval recovery.

\subsubsection{Evaluation of Waveform Morphology}
We apply the Pearson correlation ($\rho$) and relative root mean squared error (r\textsc{RMSE}) as the metrics for evaluating the ECG morphological reconstruction. They are defined as follows:

\begin{equation}
\begin{aligned}
      &\rho=\frac{(\mathbf{x}-\bar{x})^\text{T}(\hat{\mathbf{x}}-\bar{\hat{x}})}{\norm{\mathbf{x}-\bar{x}}_2 \norm{\hat{\mathbf{x}}-\bar{\hat{x}}}_2} ,\\
      &\text{r}\textsc{RMSE}=\frac{\norm{\mathbf{x}-\hat{\mathbf{x}}}_2}{\norm{\mathbf{x}}_2}.
\label{eqn:rho+rmse}
\end{aligned}
\end{equation}
where $\mathbf{x}$, $\hat{\mathbf{x}}$, $\bar{x}$, and $\bar{\hat{x}}$ denote the ground-truth ECG cycle, the recovered ECG cycle, and the average of all coordinates of the vectors $\mathbf{x}$ and $\hat{\mathbf{x}}$, respectively.

\subsubsection{Evaluation of Time Interval Recovery}
Three important ECG interval parameters are studied in this work, including the PR interval, the QRS duration, and the QT interval. Normally, the PR interval lasts 0.12-0.20 seconds, which begins from the onset of the P wave and ends at the beginning of the QRS complex. We use the segment from P point to R point of ECG as the approximated PR interval in this paper. A prolonged PR interval can indicate the possibility of first-degree heart blockage~\cite{hampton2019ecg}. The duration of the QRS complex is normally 0.12 seconds or less, for ventricular depolarization. A prolonged QRS complex indicates impaired conduction within the ventricles. The QT interval is from the onset of the QRS complex to the end of the T wave, which is normally less than 0.48 seconds. A prolonged QT interval may lead to ventricular tachycardia~\cite{hampton2019ecg}.


We apply a combination of several established algorithms~\cite{sedghamiz2018biosigkit, pan1985real, sedghamiz2015unsupervised} to detect the major fiducial points of both the ground-truth ECG and the reconstructed ECG to obtain the above-mentioned interval parameters. We apply the mean absolute error (MAE) in Eq.~\eqref{eqn:mae + rmae} to evaluate the time recovery accuracy:
\begin{equation}
\begin{aligned}
      & \textsc{MAE} = \frac{1}{N} \sum_{n=1}^N |L_{rec} - L_{ref}|.
\label{eqn:mae + rmae}
\end{aligned}
\end{equation}
where the $L_{rec}$ and $L_{ref}$ are the interval length (in seconds) of the reconstructed ECG and ground-truth ECG signals, respectively, and N is total number of cycles for evaluation. 


\begin{figure*}[!t]
\centering
\includegraphics[width = 7.0in]{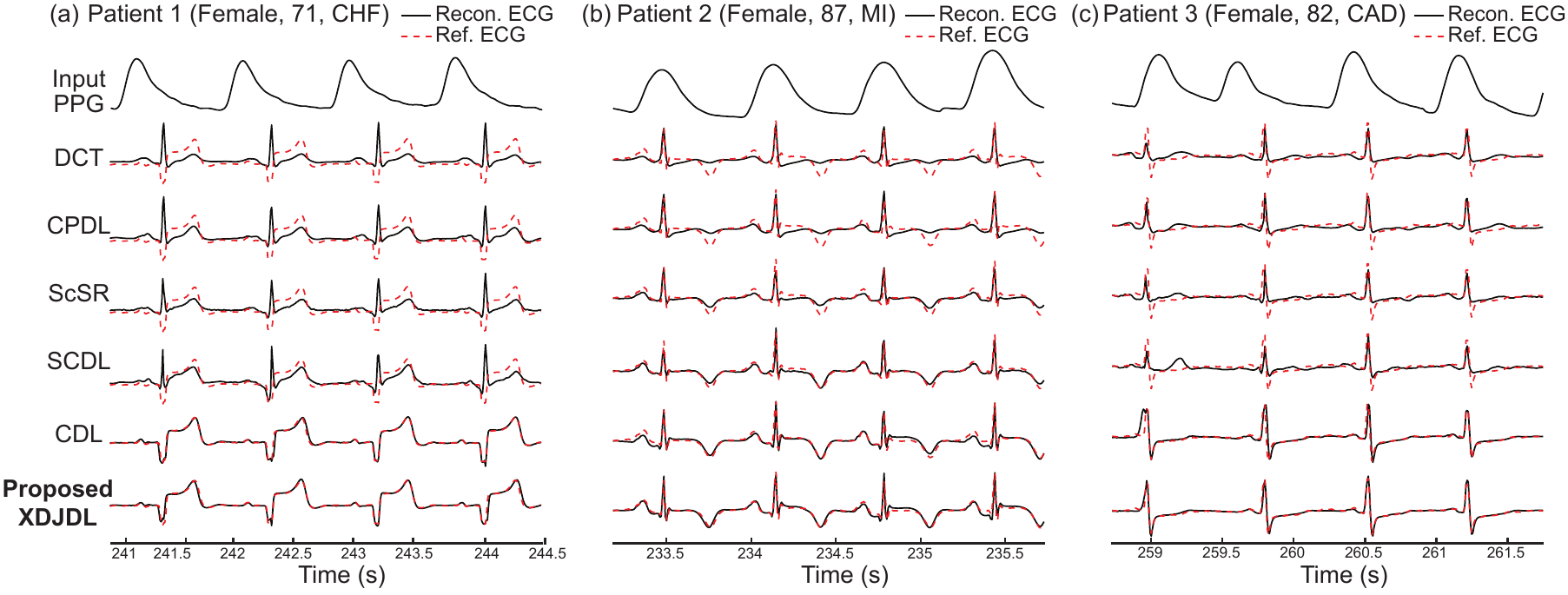}
\caption{Qualitative comparison of the ECG signals inferred by different approaches. Examples are from (a) a 71-year-old female with congestive heart failure, (b) an 87-year-old female with myocardial infarction, and (c) an 82-year-old female with coronary artery disease. From top to bottom: the input PPG signal from which the ECG is inferred in subject-independent mode, results by DCT method~\cite{2019arXiv190410481Z}, CPDL\cite{li2018discriminative}, ScSR~\cite{yang2010image}, SCDL~\cite{wang2012semi}, CDL~\cite{xu2014coupled}, and our proposed XDJDL.}
\label{fig::overall result waveform}
\end{figure*}

\subsection{Overall Morphological Reconstruction}\label{overall morph. comp.}

We compare our proposed XDJDL method with the state-of-the-art in ECG reconstruction from PPG, which used DCT based method~\cite{2019arXiv190410481Z}. In addition, we apply several representative and state-of-the-art models of coupled or semi-coupled dictionary learning, including CPDL\cite{li2018discriminative}, ScSR~\cite{yang2010image}, SCDL~\cite{wang2012semi}, and CDL~\cite{xu2014coupled}, to compare with the proposed XDJDL method on the ECG reconstruction task. The codes for the prior art are downloaded from the respective authors' websites.

To make a fair comparison, we evaluate the DCT based reconstruction system in the \textit{subject-independent} training mode where a linear transform $\mathbf{W}_{\mathrm{DCT}}$ is learned using training data from all patients. The normalized PPG/ECG cycle length is chosen as $d = 300$. For XDJDL, the dictionary size for ECG cycles is $k_e = 320$, and the dictionary size for PPG cycles is $k_p = 9000$. The sparsity parameters are set to be $t_e = 10~\text{and}~t_p = 10$. The weights for regularization terms are $\alpha = 1$ and $\beta = 1$. For other dictionary learning models, we have also done the grid-search for hyperparameter tuning to achieve the best performance. We split the data from each patient into training and test sets, and the training data ratio is~$80 \%$. 

\begin{table}[!t]
\centering
\caption{Quantitative performance comparison for ECG waveform inference}
\label{tab::all dictionary models comp}
\begin{tabular}{@{}ccccccc@{}}\toprule
\multirow{2}{*}{\begin{tabular}[c]{@{}c@{}}Reconstruction\\ Scheme\end{tabular}}~~~~~ & \multicolumn{3}{c}{\textsc{$\rho$}} & \multicolumn{3}{c}{r\textsc{RMSE}}\\
\cmidrule(r){2-4} \cmidrule(l){5-7} 
& $\hat{\mu}$& $\hat{\sigma}$& $med$ & $\hat{\mu}$& $\hat{\sigma}$ & $med$\\\midrule
DCT~\cite{2019arXiv190410481Z}  ~  & 0.71 & 0.31 & 0.83 & 0.67 & 0.26 & 0.60 \\
CPDL~\cite{li2018discriminative} ~  & 0.74 & 0.31 & 0.85 & 0.63 & 0.35 & 0.56 \\
ScSR~\cite{yang2010image}~ & 0.82 & 0.23   & 0.89  & 0.54 & 0.21  &  0.52  \\
SCDL~\cite{wang2012semi} ~ & 0.83 & 0.21 & 0.89 & 0.52 & 0.22 & 0.49 \\
CDL~\cite{xu2014coupled}  ~  & 0.85 & 0.25 & 0.95 & 0.49 & 0.51 & 0.34 \\
XDJDL (\textbf{proposed})~~~ & \textbf{0.88} & 0.23 & \textbf{0.96} & \textbf{0.39} & 0.31 & \textbf{0.29}\\\bottomrule
\end{tabular}
\end{table}

Table~\ref{tab::all dictionary models comp} shows the quantitative comparison of the ECG morphological reconstruction performance. From the statistics of the sample mean, standard deviation, and median of $\rho$ and $\text{r}\textsc{RMSE}$, we can see that our proposed XDJDL method outperforms both the DCT based algorithm and other representative coupled/semi-coupled dictionary learning models. Specifically, the average $\mathrm{r}\textsc{RMSE}$ is reduced from $0.49$ to $0.39$, or $20.4\%$ lower than CDL\cite{xu2014coupled}, which is the second-best among all competing models.

In Fig.~\ref{fig::overall result waveform}, we present visualization examples of ECG waveform reconstruction using all the competing models and our proposed XDJDL model. The three patients have different types of disease diagnosis. We observe that even though the waveform variances between the PPGs are relatively smaller than those between the ECGs, our proposed XDJDL method can recover most of the details well in the ECG signal from the PPG signal, suggesting that our method has preserved the intrinsic relation between the atoms from PPG and ECG dictionary pair. In particular, for the second-best CDL~\cite{xu2014coupled} method that can reconstruct the overall shape of ECG cycles reasonably well, it has glitches in recovering the details, such as the P wave of the first and last cycle of Patient 2 and the QRS complex of the first cycle of Patient 3.

\begin{table}[!t]
\centering
\caption{Comparison of ECG signal inference among XDJDL, LC1-XDJDL, and LC2-XDJDL methods}
\label{tab::comp with label consistency involved}
\begin{tabular}{@{}ccccccc@{}}\toprule
\multirow{2}{*}{\begin{tabular}[c]{@{}c@{}}Reconstruction\\ Scheme\end{tabular}} & \multicolumn{3}{c}{\textsc{$\rho$}} & \multicolumn{3}{c}{r\textsc{RMSE}}\\
\cmidrule(r){2-4} \cmidrule(l){5-7} 
& $\hat{\mu}$& $\hat{\sigma}$& $med$ & $\hat{\mu}$& $\hat{\sigma}$ & $med$\\\midrule
XDJDL & 0.88 & 0.23 & 0.96 & {0.39} & 0.31 & 0.29\\
LC1-XDJDL& 0.90 & 0.20 & 0.96 & {0.36} & 0.28 & 0.27\\
LC2-XDJDL & \textbf{0.92} & 0.17 & \textbf{0.97} & \textbf{0.33} & 0.25 & \textbf{0.26}\\\bottomrule
\end{tabular}
\end{table}

\begin{figure}[!t]
\centering
\includegraphics[width = 3.2in]{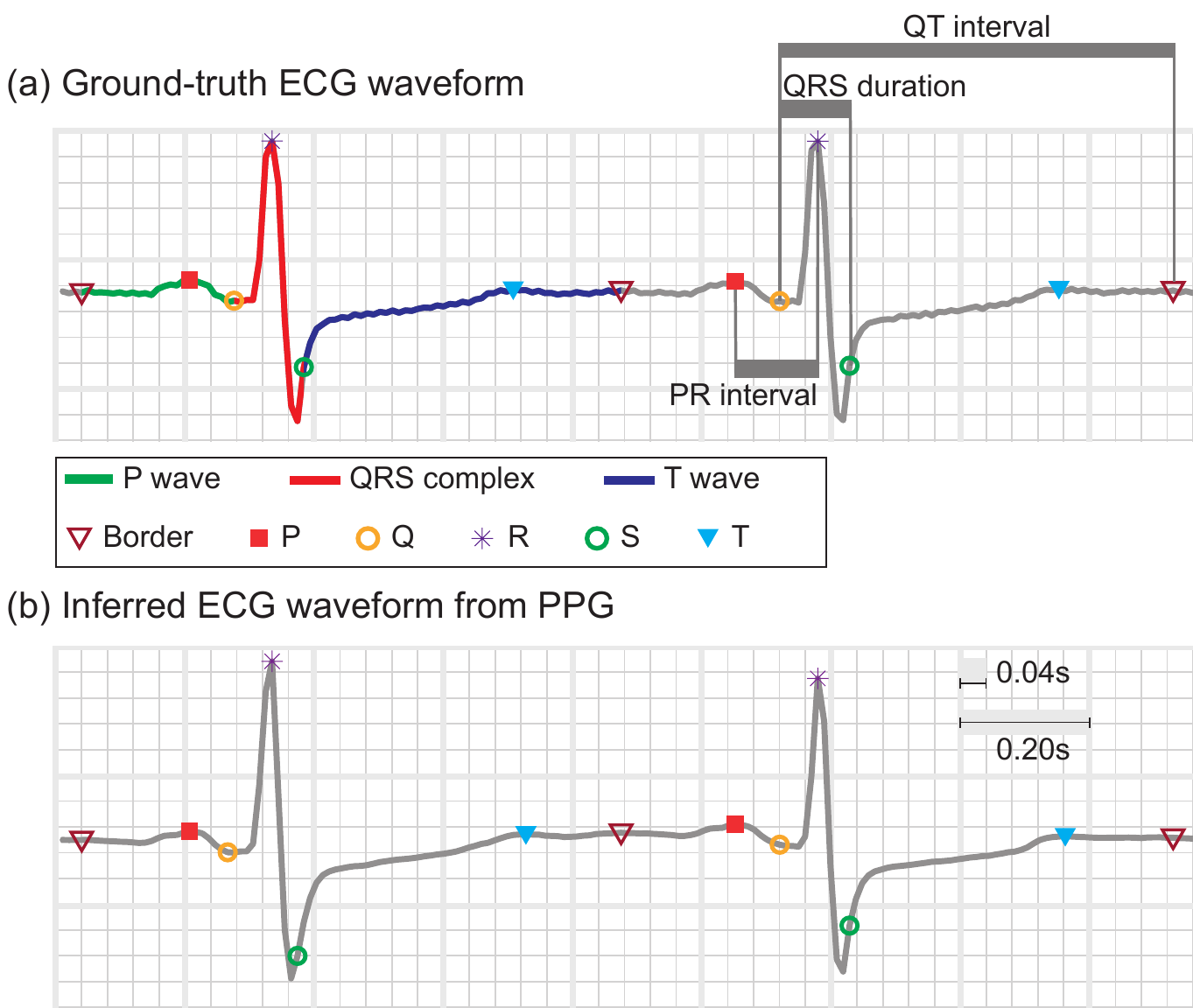}
\caption{(a) shows two cycles of the reference ECG signal and (b) shows two cycles of the inferred ECG signal. In the first cycle of (a), the green curve represents the P wave, the red curve is the QRS complex, and the dark blue curve shows the T wave. The PR interval, QRS duration, and the QT interval are all labeled in the second cycle of (a).}
\label{fig::wave seg}
\end{figure}

When the cycle-wise disease information is available, we can apply the proposed label-consistent XDJDL (LC-XDJDL) model from Section~\ref{lc-XDJDL} to leverage the label information for more accurate monitoring of ECG from the PPG signal. We consider the following scenarios: 1) For cases where the disease information is not directly provided in the test phase, we can first predict that from the PPG signals. Here, we have trained an SVM classifier for the PPG multi-class disease classification and chosen the best hyperparameters with a five-fold cross-validation method. The classification accuracy for the PPG test set reaches $92\%$. We denote the corresponding label-consistent model as LC1-XDJDL. It will take the predicted labels to build the discriminative representation matrix $\mathbf{Q}$. 2) When we have the ground-truth disease labels in the test phase, we can leverage that disease information directly as matrix $\mathbf{Q}$ and the corresponding model is named LC2-XDJDL. 

We list the comparison of ECG reconstruction performance using the XDJDL, LC1-XDJDL, and LC2-XDJDL models in Table~\ref{tab::comp with label consistency involved}. On average, the Pearson coefficient improves from 0.88 to 0.90 with the predicted label information, and to 0.92 with the ground-truth disease type as input. The improvement in terms of the rRMSE is also consistent with the Pearson coefficient. In addition to the reconstruction performance improvement, the label-consistent mapping that relates the PPG sparse codes to disease type in LC-XDJDL helps us understand the role of PPG in diagnosis with rich ECG knowledge base.

\subsection{Subwave Morphological Reconstruction} \label{subwaves morph. comp.}
In the above subsection, we have shown that our proposed XDJDL outperforms the DCT model and other representative dictionary learning models, and its performance can be better if the disease label (LC-XDJDL) can be utilized for ECG reconstruction and monitoring.

In this subsection, we zoom into the reconstruction performance of the subwave of ECG cycles using XDJDL and LC-XDJDL methods. Because each subwave refers to different atrial and ventricular depolarization and re-polarization activities, by zooming in, we can have a better idea of how our methods behave on the inference for different phases of the heart activities. A combination of the ECG major point detection algorithms~\cite{sedghamiz2018biosigkit, pan1985real, sedghamiz2015unsupervised} is used to locate P/Q/R/S/T points of ECG waveform, which helps segment the ECG cycle into subwaves for the evaluation of morphological reconstruction . 

\begin{table}[!t]
\centering
\caption{Comparison of subwave reconstructions in mean of $\rho$ and $\mathrm{r}\textsc{RMSE}$.}
\label{tab::comp with subwave morphology}

\begin{tabular}{@{}ccccccc@{}}
\toprule
\multirow{2}{*}{\begin{tabular}[c]{@{}c@{}}Reconstruction\\ Scheme\end{tabular}} & \multicolumn{3}{c}{$\bar{\rho}$}  & \multicolumn{3}{c}{$\overline{\mathrm{r}\textsc{RMSE}}$} \\ \cmidrule(lr){2-4} \cmidrule(l){5-7} & \begin{tabular}[c]{@{}c@{}}P \\ wave\end{tabular} & \begin{tabular}[c]{@{}c@{}}QRS \\ complex\end{tabular} & \begin{tabular}[c]{@{}c@{}}T \\ wave\end{tabular} &\begin{tabular}[c]{@{}c@{}}P \\ wave\end{tabular} & \begin{tabular}[c]{@{}c@{}}QRS\\ complex\end{tabular} & \begin{tabular}[c]{@{}c@{}}T \\ wave\end{tabular} \\ \midrule
XDJDL & 0.81   & 0.92   & 0.84   &  0.53   & 0.33  & 0.41   \\
LC1-XDJDL & 0.83   & 0.93  & 0.86     & 0.49   & 0.30   & 0.37   \\
LC2-XDJD & 0.86   & 0.94  & 0.89     & 0.45   & 0.28 & 0.34   \\ \bottomrule
\end{tabular}
\end{table}

\begin{figure}[!t]
\centering
\includegraphics[width = 3in]{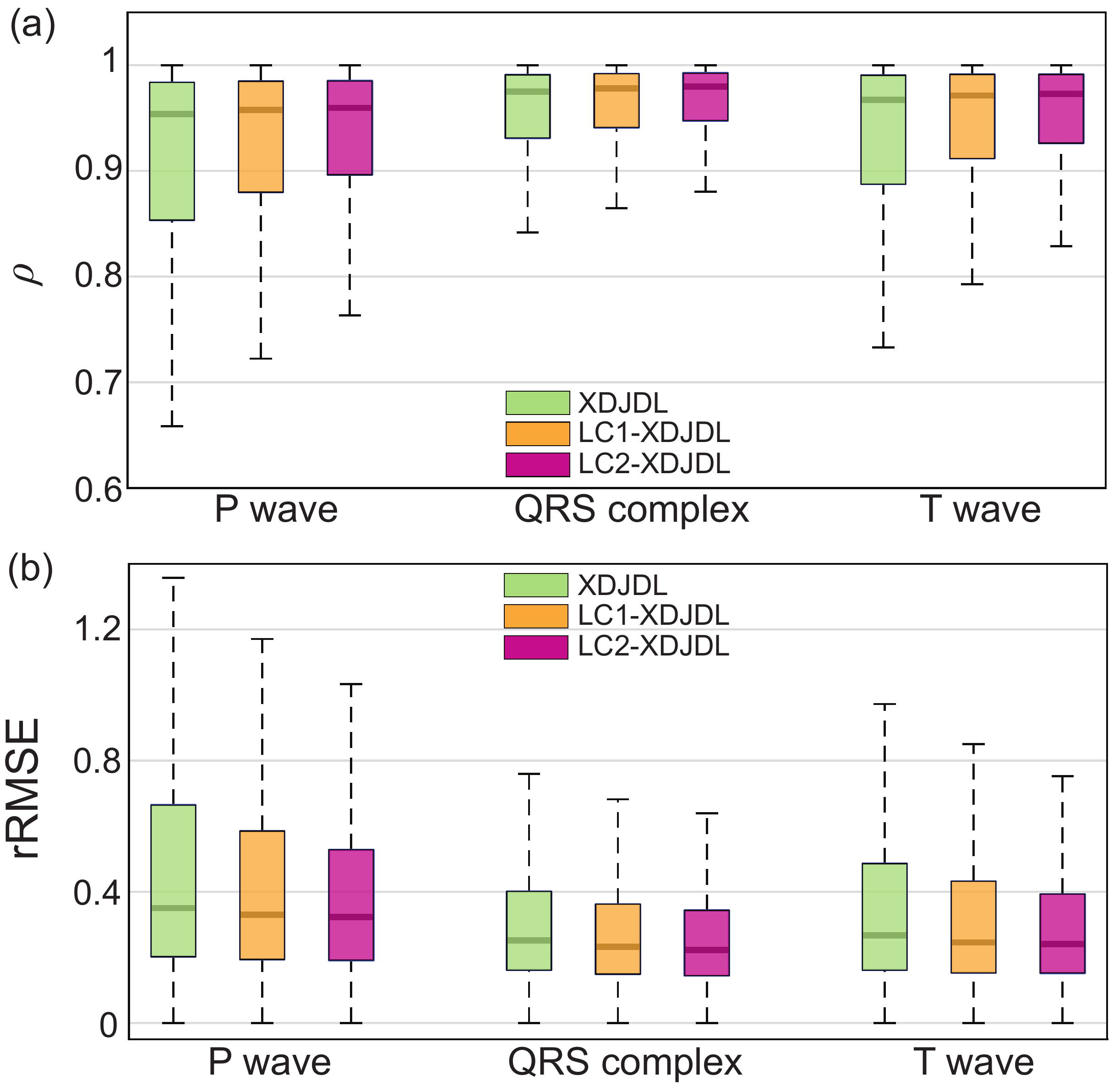}
\caption{Comparison of subwave reconstruction performance across XDJDL, LC1-XDJDL, and LC2-XDJDL models. The statistics of (a) Pearson coefficient $\rho$ and (b) r\textsc{RMSE} are summarized using the boxplots.}
\label{fig::wave seg boxplot}
\end{figure}

Fig.~\ref{fig::wave seg} shows an example of the major points detection results on two cycles of the reference ECG (Fig.~\ref{fig::wave seg}(a)) and the reconstructed ECG (Fig.~\ref{fig::wave seg}(b)) from a patient with coronary artery disease. In this example, we observe that the locations of the detected major points in both signals are close, indicating a good reconstruction of the ECG waveform. We empirically separate the adjacent ECG cycles at a point that splits the neighboring R-R peaks at the ratio of six to four. After that, a complete ECG cycle is divided into three subwaves, including the P wave that starts from the border point on the left of the ECG cycle and ends at the Q point, the QRS complex from Q to S point, and the T wave from the S point to the right border point. Only a very small portion of reference and reconstructed ECG cycle pairs cannot be detected with a consistent set of fiducial points. The overall number of effective cycles for subwave evaluation is around $92\%$ out of all test cycles, and those effective cycles only have a slightly improved Pearson coefficient ($1\%$ on average) compared to the original test dataset.

Table~\ref{tab::comp with subwave morphology} lists the reconstruction performance on the three subwaves of the ECG cycle in terms of the mean of Pearson coefficient and rRMSE using XDJDL, LC1-XDJDL, and LC2-XDJDL models. The comparison of results across models is consistent with the results of the overall comparison in Table~\ref{tab::comp with label consistency involved}. We also observe that the reconstruction for the QRS complex is better than that for the T wave, which is better than that for the P wave. The mean Pearson coefficient of the QRS complex by LC2-XDJDL is 0.94, higher than the overall cycle reconstruction of 0.92, while that of the T wave is slightly lower than the overall performance with the mean Pearson coefficient as 0.89 and that of the P wave is 0.86.

In addition to the mean of Pearson coefficient and rRMSE, Fig.~\ref{fig::wave seg boxplot} shows the comparison of the statistics of Pearson coefficient and rRMSE in boxplots for the three subwaves of ECG so that we can see the overall result distribution of the two metrics. We observe that the medians of $\rho$ and rRMSE for each of the three subwaves are very similar across the proposed models. Specifically, the medians of $\rho$ of P wave are 0.95, 0.96 and 0.96, respectively, those of QRS complex are all 0.98, and those of T wave are all 0.97; the medians of rRMSE of P wave are 0.35, 0.33, 0.32, those of QRS complex are 0.25, 0.23, 0.22, and those of the T wave are 0.27, 0.25, and 0.24, respectively. Analysis of these the boxplots suggests that our proposed models can preserve the relation between PPG and QRS complex well. The overall reconstruction performance can be improved if the relations between PPG and P and T waves are better learned.

\subsection{Time Interval Recovery} 
\label{subsec:: time interval results}

In addition to the morphological reconstruction evaluation, we evaluate whether the time intervals are well preserved. The labeling of those intervals is shown in Fig.~\ref{fig::wave seg}.

\begin{table}[!t]
\centering
\caption{Comparison of timing interval recovery accuracy in MAE.}
\label{tab::comp with sub timing interval}
\begin{tabular}{@{}ccccccc@{}}
\toprule
\multirow{2}{*}{\begin{tabular}[c]{@{}c@{}}Reconstruction\\ Scheme\end{tabular}} & \multicolumn{3}{c}{Mean (in seconds)}&\multicolumn{3}{c}{MAE (in seconds)}\\ \cmidrule(lr){2-4} \cmidrule(lr){5-7} & PR   & QRS    & QT & PR   & QRS    & QT\\ \midrule
XDJDL& 0.164 & 0.115& 0.331& 0.030  & 0.012  & 0.030 \\
LC1-XDJDL & 0.166&0.116 &0.331 & 0.026  & 0.011  & 0.027\\
LC2-XDJDL & 0.167& 0.115 & 0.331 & 0.025  & 0.010  & 0.025\\ \bottomrule
Reference & 0.172& 0.113 & 0.328& -  & -  & - \\\bottomrule
\end{tabular}
\end{table} 

From column~2-4 in Table~\ref{tab::comp with sub timing interval}, we can compare the average of the reconstructed intervals and the reference intervals. For PR intervals, the difference between the reconstructed and reference is approximately 4\%; for QRS durations, such difference is within 3\%; and for QT intervals, the difference is less than 1\%. This suggests that, on average, the timing information of the intervals is preserved well. From column~5-7 in Table~\ref{tab::comp with sub timing interval}, we also notice that the MAEs of the PR interval are 0.030s, 0.026s, and 0.025s using XDJDL, LC1-XDJDL, and LC2-XDJDL models, respectively. The relatively large errors in the timing of PR interval recovery is consistent with the result of P wave reconstruction performance shown in Section~\ref{subwaves morph. comp.}. Nevertheless, the MAE of the timing for the QRS complex is around 11ms, which is just a quarter of the smallest grid on the conventional hand copy of ECG recorders (40 ms) and is negligible given the sampling rate (125 Hz) of the ECG signal in the MIMIC III dataset. The MAE of the QT interval is around 27ms, which is less than three-quarters of the smallest grid on ECG graph paper and is around $8\%$ of the QT interval (0.331s).

\section{Discussions}
\subsection{Result Using PPG-based Segmentation Scheme}

\begin{table}[!t]
\centering
\caption{Quantitative Comparison of different segmentation schemes}
\label{tab::discussion1}
\begin{tabular}{@{}ccccccc@{}}\toprule
\multirow{2}{*}{\begin{tabular}[c]{@{}c@{}}Reconstruction\\ Scheme\end{tabular}}~~~~~ & \multicolumn{3}{c}{\textsc{$\rho$}} & \multicolumn{3}{c}{r\textsc{RMSE}}\\
\cmidrule(r){2-4} \cmidrule(l){5-7} 
& $\hat{\mu}$& $\hat{\sigma}$& $med$ & $\hat{\mu}$& $\hat{\sigma}$ & $med$\\\midrule
XDJDL (O2O)  & 0.80 & 0.24 & 0.88 & 0.55 & 0.32 & 0.48\\
XDJDL (R2R)  & 0.88 & 0.23 & 0.96 & 0.39 & 0.31 & 0.29\\\bottomrule
\end{tabular}
\end{table}

In Section~\ref{sec:exper}, we have evaluated our proposed models based on the assumption that the cycle information from ECG signals is available to separate the ECG/PPG time-series signals into training and test cycles. But in practice, we may not have the cycle information of ECG from the test data. Thus, we consider such realistic scenarios of reconstructing the ECG only from the ``estimated cycles" of PPG. Here we train on the same training data in Section~\ref{sec:exper}. For the test data, we use the PPG onsets as the segmentation points rather than the R peaks of ECG signals. The PPG onsets are used for segmentation rather than the PPG peaks because of the underlying physiological meaning as we have mentioned in Section~\ref{subsec::preproc}. For ease of notation, we use `O2O' to denote the segmentation scheme based on PPG onsets, and `R2R' to denote the segmentation scheme based on R peaks of ECG.

Due to the discrepancy between the detected locations of PPG onset and R peak of ECG from the same cycle, the ``estimated PPG cycles'' using O2O scheme slightly vary from the test PPG cycles which are segmented by R2R. To single out the contribution to the ECG reconstruction error due to the discrepancy in the waveform shape rather than the misalignment of the ECG peaks, we evaluate O2O after compensating the time offset between the reconstructed ECG and original ECG signals. This is done by shifting each reconstructed ECG cycle in time so that the original and reconstructed ECG signals are matched according to their R peaks. The result comparison is shown in Table~\ref{tab::discussion1}. On average, the Pearson coefficient drops from 0.88 to 0.80, and the rRMSE rises from 0.39 to 0.55. 


\subsection{Performance of Leave-One-Out Experiment}

As a proof of concept and considering the current moderate amount of the available data, we have so far split each patient's data into training and test sets. This corresponds to the trending of ``precision medicine" to tailor the healthcare practice to individual patients. Meanwhile, we are curious how the system would behave if the test patient is never seen in the training phase, corresponding to the situation of training models for the whole population or patient groups categorized by gender, age, race or other ways. We will examine this through leave-one-out experiments. 

We apply a pre-clustering process based on the ECG data to select a sub-group of patients with similar ECG features for the leave-one-out experiment. First, we reduce the dimension of the ECG cycles by principal component analysis (PCA), and then we use K-means to cluster the ECG features after PCA. Based on the clustered ECG features, we select the largest cluster of ECGs from 19 patients. The mean Pearson coefficient for the leave-one-out experiment on the 19 patients is 0.74 (std: 0.15, median: 0.77).

From this preliminary result, we can see that as expected, the leave-one-out experiment is a more challenging case given the large variability of ECG data morphologies of ICU patients and the limited number of patients in the collected dataset. Based on the results in Section~\ref{overall morph. comp.}, we see the encouraging capability of recovering large variations in ECG from relatively small variations in PPG across cycles and patient populations. This suggests a strong potential of predicting ECG from PPG of unseen patients through further research and larger data collection. In our follow-up work, we are considering an improved problem definition and data collection procedure to improve the generalization capability of the learned system.

\subsection{Towards Explainable AI}

Our proposed XDJDL and LC-XDJDL models accomplished to infer the ECG based on PPG by leveraging the biomedical and statistical relationship between the signals. This is an initial effort to demonstrate a potential benefit from our ``explainable'' AI, rather than black-box data-driven AI, to provide the the more user-friendly PPG measurements inferred ECG data for the medical professionals to interpret and offer medical insights. Our framework also helps transfer the rich ECG knowledge base from decades of medical practice to augment the PPG diagnosis for public health. 

Given the challenge of making the ECG inference more accurate for an unseen group of subjects, e.g., by age, gender, or other medical and health condition, we are extending our current work with a neural network to further enrich the representation and learn the relation when sufficient data is available. Our ongoing efforts have been focused on both developing a data collection pipeline for more diversity and coverage of training data and exploring an explainable generative model with strong expressive power to improve the generalization performance. With the step-by-step capturing of the complex models by utilizing the biomedical, statistical, and the physical meanings, as well as harnessing the power of the data, we aim to provide explainable AI with our ongoing efforts.

\section{Conclusions}\label{sec:conclusion}
We have proposed a cross-domain joint dictionary learning (XDJDL) framework and the extended label-consistent XDJDL (LC-XDJDL) model for ECG waveform inference from the PPG signal to facilitate continuous cardiac monitoring. The promising experimental results validate that our proposed models can learn the relation between PPG and ECG well. From the analysis for subwave reconstruction and timing of interval recovery, we observe that we can preserve the QRS complex, QRS duration, and the QT interval very well, which is essential in the ECG monitoring and gaining more PPG-based diagnosis knowledge. This work reveals the potential of using explainable AI to realize the long-term and user-friendly ECG screening from the PPG signals that we can acquire from daily use of wrist watch or fingertip pulse oximeter. Our ongoing work and future endeavor will strive to improve the accuracy and generalization capability of the ECG waveform inference from PPG while maintaining sound interpretability.


\bibliographystyle{IEEEtran}
\bibliography{refs}

\end{document}